\documentclass[pss,fleqn]{w-art}
\usepackage{times}
\usepackage{w-thm}
\usepackage{epsfig} 
\usepackage[]{graphicx}
\begin{document}
 \renewcommand{\copyrightyear}{2004}
\DOIsuffix{theDOIsuffix}
\Volume{XX}
\Issue{1}
\Month{01}
\Year{2004}
\pagespan{1}{}
\Receiveddate{21 October 2004}           



\title{Underetching from simple stochastic etching kinetics
}


\author[Jens Christian Claussen]{Jens Christian Claussen\footnote{Corresponding
     author: e-mail: {\sf claussen@theo-physik.uni-kiel.de}, Phone:
+49\,431\,880\,4096,
     Fax: 
+49\,431\,880\,4094}\inst{1}} \address[\inst{1}]{Institut for Theoretical
Physics, University Kiel, Leibnizstr.\ 15, 24098 Kiel, Germany}
\author[J\"urgen Carstensen]{J\"urgen Carstensen\footnote{e-mail: {\sf
jc@tf.uni-kiel.de}}\inst{2}}
\address[\inst{2}]{Chair of General Materials Science, University Kiel, 
Kaiserstr.\ 2, 24143 Kiel, Germany}
\begin{abstract}
The morphological richness of electrochemical semiconductor 
etching is not sufficiently counterparted yet by theoretical
modeling. 
This paper investigates a minimal version of the Current-Burst 
model with Aging of F\"oll and
Carstensen and demonstrates for a restricted geometry
that the Aging concept is essential for underetching,
or cavity generation.
If the influence of Aging is neglected, the dynamics 
reduces to a Random Etching Model similar to
the Random Deposition model.
This computer {\sl gedanken experiment} demonstrates that 
the stochastic dynamics with ageing-dependent 
kinetic reaction probabilities accounts for the 
different etching morphologies compared to
those obtained in surface roughening and related systems.
\end{abstract}
\maketitle                   






The electrochemical semiconductor-electrolyte interface 
shows various types of highly nonlinear and non-equilibrium 
phenomena. 
Various reasons come into play: 
First, any chemical reaction 
---~if not well stirred~---
is a nonlinear and spatial dynamical system 
as the reaction rates are a product of at least
two reactand densities.
Second, most etching systems, especially in production
systems, are driven with a high current density,
resulting in a far-from-equilibrium situation.
Third, local clustering of etching activity is
energetically favorable to a homogeneous density,
resulting in cooperative phenomena and synchronization.
Fourth, hydrodynamic and diffusion limitations
delimit reaction rates.
The interplay between nonlinear kinetics
and e.g.\ diffusion limitation gives rise to
rich spatio-temporal pattern formation.
Reaction-diffusion systems have been studied widely
and model various inhomogeneous modes of pattern
formation as travelling waves, solitonic structures,
and spiral waves, and been widely applied to
catalytic reactions. 
Yet the morphology of the surface remains unchanged.
For etching, or corrosion, the surface atoms are
not inert as in catalysis, but themselves take part
in the reaction dynamics.
In addition that morphology is modified according to
local concentrations, also the morphology influences the
concentrations.

\paragraph*{Current-Burst Model with Aging}
The Current-Burst (CB) model developed by F\"oll and
Carstensen 
\cite{foellreview,foellnewview}
explains qualitatively a large variety of 
semiconductor etching experiments
from a stochastic nonequilibrium dynamics based on 
very few assumptions.
Due to the large range of scales involved in space and time,
neither ab initio methods nor a full 3D simulation 
with CB size evolution are computationally feasible
if one wants to explain e.g.\
branching morphologies 
and their open-loop control suppression
\cite{claussen03}
or even fractal structures
\cite{marcPHD,claussen03comphys,claussen03physcon}.
The qualitative understanding has developed 
quite far to a detailed understanding
of the different morphologies in different types of semiconductors
(n/p type, Si, Ge, II/V-compounds), different etchants
(HF, organic) and different parameters (front/backside illumination,
doping level, temperature, current amplitude and waveform).

The Current Burst Model can be defined as follows.
Etching, i.e.\ the dissolution of semiconductor atoms,
occurs only within {\sl bursting} events localized both in
time and space, a typical size can be $10^3$ atoms.
The CB is an irreversible process far from equilibrium
with initially high energy dissipation density;
local 
high electrical field strength
 induces cracks within a characteristic radius,
along those, atoms are dissoluted on a short time scale.

New etching events are initialized where surface passivation
e.g.\ 
by hydrogen is absent, i.e.\ unsaturated binding valences 
are present. 
Their density is highest immediately after the etching event,
differs for the different surface orientations 
in which a disordered surface dominatly facets 
(e.g.\ $<100>$ and $<111>$ in Si)
and decreases exponentially with time,
with different time constants for the surface orientations.
As a consequence, the initialization density
of events, which depends linearly on the
density of available bonds, itself decreases exponentially with 
time, until it saturates to a very low 
probability density similar to that of the inert surface. 

In the stochastic picture, these kinetic reaction
rates take the role of reaction probabilities.
Hereby the corresponding kinetic Monte Carlo model
is defined; 
the experimental default setup of 
galvanostatic etching is
straightforwardly modeled by a normalization of
the field of local reaction probabilities
over all surface elements.

In a fully detailed model, 
the electric potential within the bulk, 
diffusion limitations, the electrochemic double layer etc.\
should be taken into account.
This is neglected within the model presented here;
not aiming at an exact morphology prediction,
but demonstrating the difference 
to reaction kinetics as known in surface growth 
and surface roughening.

\noindent


\paragraph*{Minimal CB model in a 2D cross section}
Experimentally pore
morphologies are analyzed by breaking the wafer and investigating the
cleavage plane with a light or electron microscope
--- no in situ imaging technique is applicable
during the etching.
Thus experimental morphologies are 
available only for those cross-sections.
Consequently, the three-dimensional arrangement of pores
may be neglected in a first simplified model,
i.e.\  a 2D cross section is studied.
Once the 2D system is fully understood, the
more complicated and computationally costly 3D case
can be addressed also.

The minimal model is defined on a orthogonal
lattice, and consists of cells or
plaquettes (of the size ranging from one atom
to one CB) being occupied (1) or empty (0).
Every bond between a 1 cell and a 0 cell corresponds to a surface 
element.

The minimal model is implemented by a 
kinetic Monte-Carlo simulation where 
each surface element has a local memory of the time 
where it became a surface element.
Then CB's are initialized according to both the
field of local reaction probabilities and 
the pre-set current density (of plaquettes per time step).
Each element that is chosen for reaction in the
Monte Carlo process initiates dissolution
of all cells within a certain radius.
In the simulations presented here, 
a minimal size of CB's is used
(of radius 1/2),\footnote{%
Nota bene {\sl this implementation} bears ``geometric artefacts'', thus 
 is not expected to predict the exact spatial structure.
 Here $<11>$ model surfaces have compared to $<10>$ model surfaces 
 a factor $\sqrt[]{2}$ higher density of surface elements,
 contrary to a factor 1 in the real system. 
 This leads to favored etching in $<11>$ direction
 contrary to the situation in Si, where the $<10>$ 
 direction is preferred in etching.
 In a refined model, this could be corrected in the model by a rescaling
 of the ratio of the passivation time constants. }
and the CB model is studied with istotropeous aging
under galvanostatic conditions.

\begin{figure}[htb]
\includegraphics[width=\textwidth]{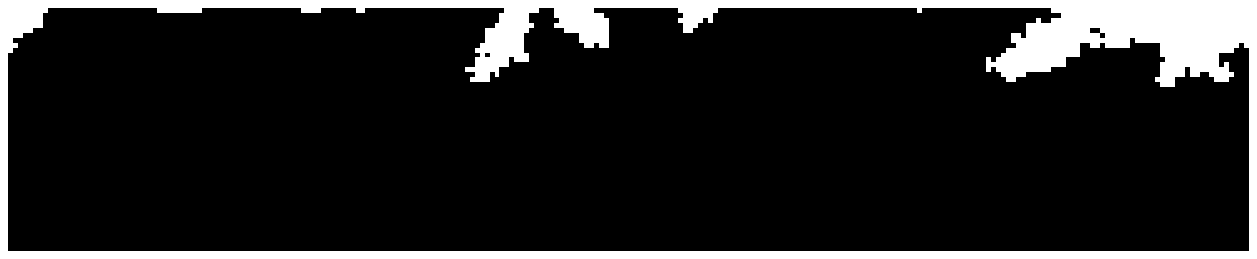}
\caption{Kinetic Monte Carlo simulation of the Current-Burst Model with Aging. 
The system shows strong underetching and readily
generates cavities. 
Here 250 cells with periodic boundary conditions are used.
\label{fig:1}}
\includegraphics[width=\textwidth]{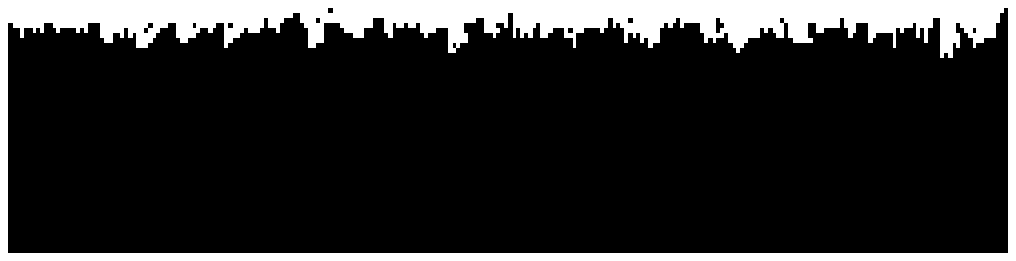}
\caption{Random Etching (or Random Aggregation) Model: 
Uncorrelated stochastic etching,
resulting in surface roughening.
Underetching is possible, but occurs rarely. \label{fig:2}}
\end{figure}

\paragraph*{Results. 1.}
The system shows 
strong underetching and readily 
generates
cavities, as shown in Fig.~\ref{fig:1}.
Due to the geometric artefact, etching in
$<11>$ direction is favored, contrary to the
situation in Si, where the  $<10>$ direction
etches dominantly.
Thus straight-wall pore geometries 
are not obtained within this model.

\paragraph*{2. Comparison to a ``Random Aggregation'' (or Etching) Model}
If all kinetic rates are set to a homogeneous and
time-independent constant, the
morphology creating effect of 
{\sl space-time correlations} of CB's due to
the aging must vanish; 
therefore only disordered etching is expected.
This effect demonstrates easily, as shown in Fig.~\ref{fig:2},
no other parameters have beein changed.

If all kinetic rates are equal, 
the model corresponds to an isotropic 
(``non-MBE'') version of Random Deposition Model, 
or ``not-diffusion-limited'' DLA.
--- 
While the RD model, the Eden model, DLA and other 
variants have been studied extensively 
(see \cite{barabasistanley} for a review
and introduction of scaling concepts in surface growth),
this (isotropic-deposition) random aggregation
(or etching) model seems to be less studied,
e.g. it should be  clarified to which universality class
it belongs and which scaling exponents describe the surface roughening in this model.

\paragraph*{Discussion and Outlook}
The model investigated shows that the locally time-dependent dynamics resulting 
from the aging concept introduces geometries dramatically different from
those known from surface roughening:
Underetching of cavities is favored, and inert sites remain
unattached for a long period.
This behavior is absent in known models of 
surface roughening.
If aging effects are switched off in the simulation,
surface roughening kinetics is restored.
This model thus provides a ``test plant'' to check 
theoretically the effect
of switched-on/off aging.
Apart from this qualitative result,
the different aging kinetics for each surface
orientation should be included as well as 
Current Burst sizes of more than one lattice
constant should be taken into account.
This will be subject of further investigation.
\begin{acknowledgement}
J.C.C.\ wants to thank Helmut F\"oll and J\"urgen Carstensen
for stimulating discussions.
\end{acknowledgement}


\begin{thebibliography}{10}
\bibitem{foellreview} H. F\"oll, 
M. Christophersen, S.\ Langa, G.\ Hasse,
Mat.\ Sci.\ Eng.\ R 39, 4, 93-141 (2002).

\bibitem{foellnewview}
 H.\ F\"oll, J.\ Carstensen, M.\ Christophersen, and G.\ Hasse,
Phys.\ Stat.\ Sol.\ (a), 182, 
7 (2000).



\bibitem{claussen03}
J.C.\ Claussen, J.\ Carstensen, M.\ Christophersen, S.\ Langa, and H.\ F\"oll (2003), 
Chaos 13 (1),
217-224.


\bibitem{marcPHD} Marc Christophersen, PhD thesis, Kiel, 2002.



\bibitem{claussen03comphys}
J.C.\ Claussen, J.\ Carstensen, M.\ Christophersen, S.\ Langa, and H.\ F\"oll (2003), 
 p. 82-87 in: Interface 
\&
Transport Dynamics, ed. H.\ Emmerich, B.\ Nestler, M.\ Schreckenberg,
 Lect.\ Notes
Comp.\  Sci.\  Eng.\
 Vol. 32,
Springer 2003.

\bibitem{claussen03physcon}
J.C.\ Claussen, J.\ Carstensen, M.\ Christophersen, S.\ Langa, and H.\ F\"oll (2003), 
in: Alexander L. Fradkov (ed.): Proceedings PHYSCON 2003, St. Petersburg,
Vol. 3, pp. 895-900 (2003).


\bibitem{barabasistanley}
A.-L.\ Barabasi and H.E.\ Stanley,
Fractal Concepts in Surface Growth,
Cambridge University Press (1995).



\end{thebibliography}
\end{document}